\documentclass[epj]{svjour}

\usepackage[numbers,sort&compress]{natbib}
\RequirePackage[colorlinks,citecolor=blue,urlcolor=blue,linkcolor=blue]{hyperref}
\usepackage{amssymb}
\usepackage{amsmath}
\usepackage{graphicx}
\usepackage{multirow}
\usepackage{hyperref}
\usepackage{natbib}
\usepackage{array}
\usepackage{float}
\usepackage{widetext}
\usepackage{flushend}
\usepackage{cuted}
\allowdisplaybreaks
    
\begin{document}

\title{Analysis of excited quark propagator effects on neutron charge form factor}

\author{X. Y. Liu\inst{1,}\inst{2,}\thanks{lxy\_gzu2005@126.com} \and A. Limphirat\inst{2,}\thanks{ayut@g.sut.ac.th} \and K. Xu\inst{2} \and D. Samart\inst{3} \and K. Khosonthongkee\inst{2} \and Y. Yan\inst{2}}

\institute{School of Mathematics and Physics, Bohai University, Jinzhou 121013, China
\and School of Physics and Center of Excellence in High Energy Physics $\&$ Astrophysics, Suranaree University of Technology, Nakhon Ratchasima 30000, Thailand
\and Department of Physics, Faculty of Science, Khon Kaen University, Khon Kaen 40002, Thailand}

\date{Received: date / Accepted: date}

\abstract{
The charge form factor and charge radius of neutron are investigated in the perturbative chiral quark model (PCQM) with considering both the ground and excited states in the quark propagator. A Cornell-like potential is extracted in accordance with the predetermined ground state quark wavefunction, and the excited quark states are derived by solving the Dirac equation with the extracted PCQM potential numerically. The study reveals that the contributions of the excited quark states are considerably influential in the charge form factor and charge radius of neutron as expected, and the total results are significantly improved and increased by nearly four times by including the excited states in the quark propagator. The theoretical PCQM results are found, including the ground and excited quark propagators, in good agreement with the recent lattice QCD values at pion mass of about 130 MeV. 
}

\maketitle

\section{\label{sec:Intro}Introduction}
The electromagnetic form factors play a unique role to investigate the nucleon internal structure and interaction. Among the four non-strange nucleon electromagnetic form factors ($G_E^p(Q^2)$, $G_M^p(Q^2)$, $G_E^n(Q^2)$ and $G_M^n(Q^2)$), the neutron charge form factor, $G_E^n(Q^2)$, is special since the $G_E^n(Q^2)$ would vanish at all momentum transfers $Q^2$ if the SU(6) spin-flavor symmetry of QCD was exact. However, non-zero experimental values of the $G_E^n(Q^2)$ have been reported~\cite{Geis:2008} and the neutron charge radius squared is $\langle r^2_E\rangle^n=-0.116\pm0.002$~$\textrm{fm}^2$~\cite{PDG:2016}. The measurement of the $G_E^n(Q^2)$ is particularly challenging due to its small value and the lack of a high-density ``pure" neutron target. To make these measurements, complex light targets like $^2H$ and $^3He$ must be used in quasi elastic scattering. The electron elastic scattering experiments for nucleon form factors have been reviewed in Ref.~\cite{Punjabi:2015}. Very recently, the lattice QCD simulations directly at the physical point become available for the $G_E^n(Q^2)$. In Ref.~\cite{Alexandrou:2017}, the lattice QCD results for the $G_E^n(Q^2)$ at pion mass of about 130 MeV are published. 

In recent years, great attention has been paid to the theoretical study on the $G_E^n(Q^2)$ and charge radius~\cite{Thomas:1981,Buchmann:1991,Lu1:1998,Lu2:1998,Cardarelli:1999,Tang:2003,Fuchs:2004,Schindler:2005,Faessler1:2006,Faessler2:2006,Faessler:2008,Glozman:1999,Rinehimer:2009,Ramalho:2011,Ramalho:2013,Wang:2009,Alexandrou:2017,Shintani:2018}. At low momentum transfer $Q^2$, the pion cloud of the nucleon is expected to play a significant role in the quantitative description of the form factors. The study in Ref.~\cite{Buchmann:1991} indicates that the $G_E^n(Q^2)$ stems mainly from the pion and gluon exchanges in the constituent quark model (CQM) while Refs.~\cite{Lu1:1998,Lu2:1998} report the importance of the correction of center-of-mass motion to the $G_E^n(Q^2)$ form factor in the relativistic quark model (RQM). In Refs.~\cite{Fuchs:2004,Schindler:2005}, the $G_E^n(Q^2)$ is investigated in the chiral perturbation theory (ChTP) under the extended on-mass-shell renormalization scheme, and the result reveals that the $G_E^n(Q^2)$ is very sensitive to higher-order contributions. Moreover, the effects of the meson cloud to the $G_E^n(Q^2)$ have been studied and estimated in various quark meson coupling models ~\cite{Glozman:1999,Rinehimer:2009,Ramalho:2011,Ramalho:2013}. These studies form a better understanding on the role of the quarks and the meson cloud dressing. 

\begin{figure}[t]
\begin{center}
\includegraphics[width=0.36\textwidth]{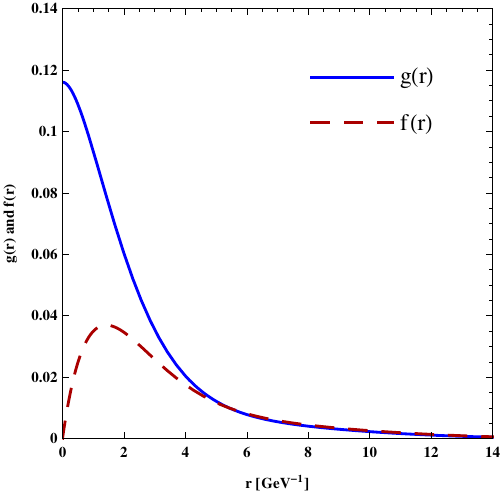}
\end{center}
\caption{Normalized radial wavefunction of the valence quarks for the upper component $g(r)$ and the lower component $f(r)$, which are determined by fitting the theoretical results of the proton charge form factor to the experimental data~\cite{Liu:2014}.}\label{fig:QWF}
\end{figure} 

\begin{figure}[b!]
\begin{center}
\includegraphics[width=0.4\textwidth]{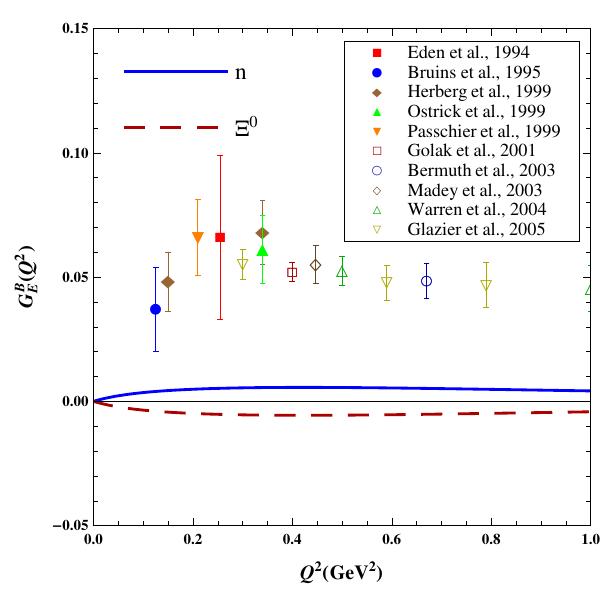}
\end{center}
\caption{Neutron charge form factor in the PCQM with the predetermined ground state quark wavefunction only~\cite{Liu:2014}.}\label{fig:GEn}
\end{figure}

The perturbative chiral quark model (PCQM) is an indispensable tool in studies of the baryon structure and properties in the low-energy particle physics~\cite{Liu:2014,Liu:2015,Lyubovitskij1:2001,Lyubovitskij3:2001,Lyubovitskij6:2002,Lyubovitskij2:2002,Pumsa-ard7:2003,Cheedket4:2004,Khosonthongkee10:2004,Dong8:2006,Dib9:2006,Faessler5:2008}. In our previous works~\cite{Liu:2014,Liu:2015}, the electromagnetic and axial form factor as well as electroweak properties of octet baryons have been evaluated in the PCQM based on the predetermined ground state quark wavefunction, which is extracted by fitting the PCQM theoretical result of the proton charge form factor $G_E^p(Q^2)$ to the experimental data~\cite{Liu:2014} as shown in Fig.~\ref{fig:QWF}. The PCQM theoretical results with the predetermined quark wavefunction are fairly consistent with the experimental data and lattice QCD values (except for neutron). That may indicate that the predetermined ground state quark wavefunction is reasonable and credible in the PCQM. However, the work failed to reproduce the experimental data of the $G_E^n(Q^2)$, seen in Fig.~\ref{fig:GEn}. As we argued in Ref.~\cite{Liu:2014}, the quark propagator is restricted to the ground-state only in the calculation, and the meson cloud solely contributes to the $G_E^n(Q^2)$ as the leading-order contribution of the 3q-core vanishes in the PCQM based on the SU(6) structure. Due to the tiny experimental values of the $G_E^n(Q^2)$, there is reason to believe that higher-order contribution is very sensitive to the $G_E^n(Q^2)$. Thus, one may propose that it is necessary to include excited-state quarks to investigate the $G_E^n(Q^2)$. More discussions and results on the neutron charge radius including the excited quark propagator may be found in Ref.~\cite{Liu:2014}. In this work, we attempt to investigate and improve the $G_E^n(Q^2)$ in the framework of PCQM with including both ground and excited states in quark propagator. It is noted that there are no further parameters to be adjusted in the present work.

The paper is organized as follows. In Sec.~\ref{sec:Potential&WF}, we extract the PCQM potential, based on the predetermined ground state quark wavefunction, and evaluate the excited quark wavefuntions. The theoretical expressions of the $G_E^n(Q^2)$ in the PCQM are listed in Sec.~\ref{sec:GEn}, and the numerical results and discussions are presented in Sec.~\ref{sec:Results}. Finally, we summarize and conclude the work in Sec.~\ref{sec:Summary}

\section{\label{sec:Potential&WF}Potential and wavefunctions}

In the framework of the PCQM, baryons are considered as the bound states of three relativistic valence quarks moving in a central potential with $V_{\textrm{eff}}(r)=S(r)+\gamma^0V(r)$, while a cloud of pseudoscalar mesons, as the sea-quark excitations, is introduced for chiral symmetry requirements, and the interactions between quarks and mesons are achieved by the nonlinear $\sigma$ model in the PCQM. The Weinberg-type Lagrangian of the PCQM under an unitary chiral rotation~\cite{Liu:2014,Liu:2015} is derived as,
\begin{eqnarray}
\mathcal{L}_0(x)&=& \bar{\psi}(x)\big[i\partial\!\!\!/-\gamma^{0}V(r)-S(r)\big]\psi(x)\nonumber\\
&&-\frac{1}{2}\Phi_i(x)\big(\Box+M_{\Phi}^2 \big) \Phi^i(x),\label{WT-0}\\
\mathcal{L}_I^W(x)&=&\frac{1}{2F}\partial_\mu\Phi_i(x)\bar{\psi}(x)\gamma^\mu\gamma^5
\lambda^i\psi(x)\label{eq:WT-int},
\end{eqnarray}
where $f_{ijk}$ are the totally antisymmetric structure constants of $SU(3)$, the pion decay constant $F=88$ MeV in the chiral limit, $\Phi_i$ are the octet meson fields,
and $\psi(x)$ is the triplet of the $u$, $d$, and $s$ quark fields taking the form

\begin{figure}[b!]
\begin{center}
\includegraphics[width=0.235\textwidth]{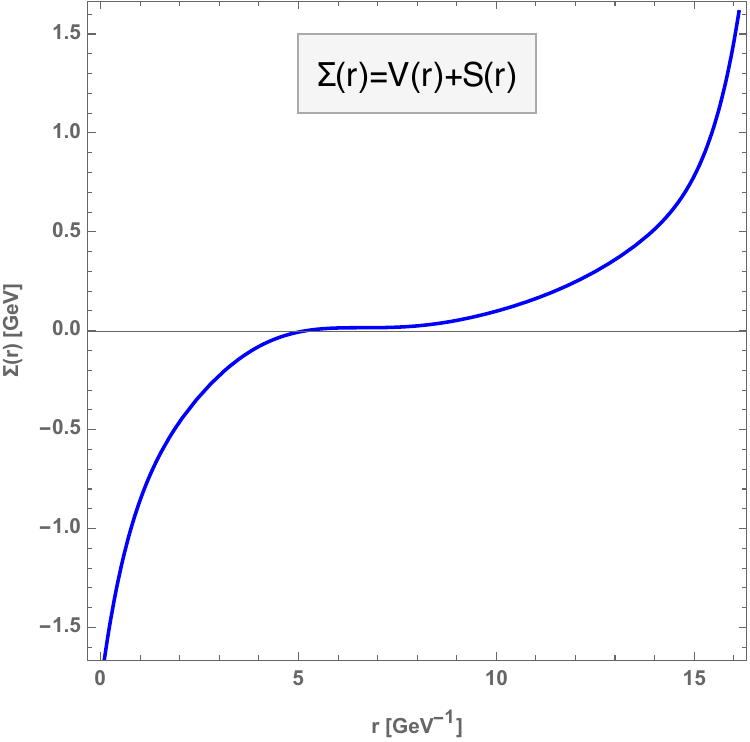}
\includegraphics[width=0.234\textwidth]{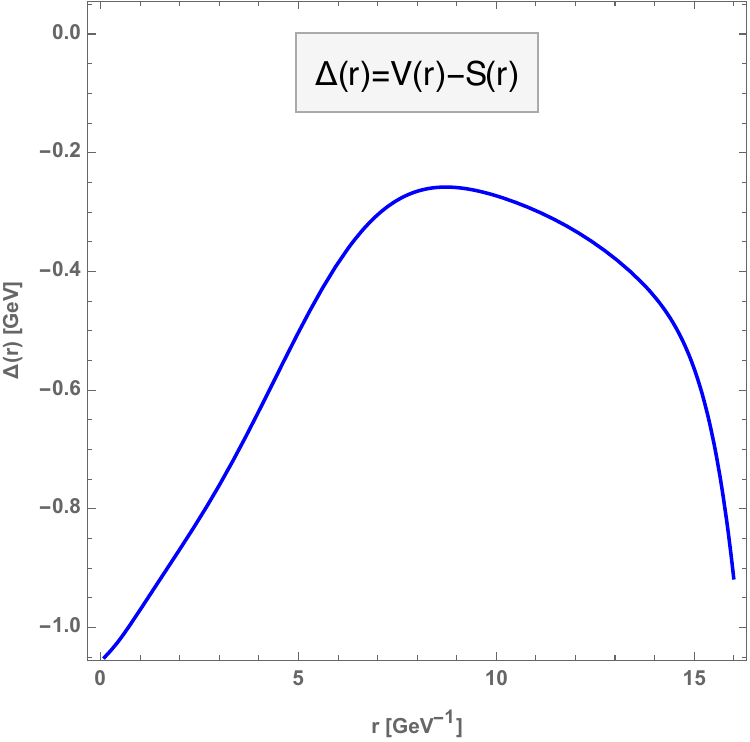}
\end{center}
\caption{\label{fig:potential}The extracted PCQM potentials based on the predetermined ground state quark wavefunction. Left panel: $\Sigma(r)$ potential, right panel: $\Delta(r)$ potential.}
\label{fig:potential}
\end{figure}

\begin{table}[b!]
\caption{\label{tab:energy} The energy levels of a single quark with $E_0=0.524$~GeV.}
\begin{tabular*}{\columnwidth}{@{\extracolsep{\fill}}cccccc}
\hline
   \multicolumn{1}{c}{Notation}&
   \multicolumn{1}{c}{{}$n$}&
   \multicolumn{1}{c}{{}$l$}&
   \multicolumn{1}{c}{{}$j$}&
   \multicolumn{1}{c}{{}$\kappa$}&
   \multicolumn{1}{c}{{}$E_\alpha$ (GeV)}\\[2pt]
\hline
   $1s_{1/2}$ & $1$ & $0$ & $1/2$ & $-1$ & $0.524$\\
   $2s_{1/2}$ & $2$ & $0$ & $1/2$ & $-1$ & $0.722$ \\
   $3s_{1/2}$ & $3$ & $0$ & $1/2$ & $-1$ & $0.935$\\
   $4s_{1/2}$ & $4$ & $0$ & $1/2$ & $-1$ & $1.122$\\
   $1p_{1/2}$ & $1$ & $1$ & $1/2$ & $\phantom{-}1$ & $0.669$\\
   $2p_{1/2}$ & $2$ & $1$ & $1/2$ & $\phantom{-}1$ & $0.847$\\
   $3p_{1/2}$ & $3$ & $1$ & $1/2$ & $\phantom{-}1$ & $1.041$\\
   $1p_{3/2}$ & $1$ & $1$ & $3/2$ & $-2$ & $0.738$\\
   $2p_{3/2}$ & $2$ & $1$ & $3/2$ & $-2$ & $0.877$\\
   $3p_{3/2}$ & $3$ & $1$ & $3/2$ & $-2$ & $1.059$\\
   $1d_{3/2}$ & $1$ & $2$ & $3/2$ & $\phantom{-}2$ & $0.805$\\
   $2d_{3/2}$ & $2$ & $2$ & $3/2$ & $\phantom{-}2$ & $1.009$\\
   $1d_{5/2}$ & $1$ & $2$ & $5/2$ & $-3$ & $0.844$\\
   $2d_{5/2}$ & $2$ & $2$ & $5/2$ & $-3$ & $1.025$\\
   $1f_{5/2}$ & $1$ & $3$ & $5/2$ & $\phantom{-}3$ & $0.920$\\
   $2f_{5/2}$ & $2$ & $3$ & $5/2$ & $\phantom{-}3$ & $1.101$\\
   $1f_{7/2}$ & $1$ & $3$ & $7/2$ & $-4$ & $0.933$\\
   $2f_{7/2}$ & $2$ & $3$ & $7/2$ & $-4$ & $1.133$\\
   $1g_{7/2}$ & $1$ & $4$ & $7/2$ & $\phantom{-}4$ & $1.022$\\
   $1g_{9/2}$ & $1$ & $4$ & $9/2$ & $-5$ & $1.028$\\
\hline
\end{tabular*}
\end{table} 

\begin{equation}
\psi(x) =\left(
\begin{array}{c}
u(x) \\
d(x) \\
s(x) \\
\end{array}
\right).
\end{equation}
The quark field $\psi(x)$ could be expanded in the form,
\begin{equation}
\psi(x)=\sum_\alpha\left(b_\alpha u_\alpha(\vec{x})\,e^{-iE_\alpha t}+d^\dagger_\alpha\upsilon_\alpha (\vec{x})e^{iE_\alpha t}\right),
\end{equation}
with $b_\alpha$ and $d^\dag_\alpha$ as the single quark
annihilation and antiquark creation operators. The set of quark $u_\alpha$ and antiquark $\nu_\beta$ wavefunctions in orbits $\alpha$ and $\beta$ is the solution of the static Dirac equation:
\begin{equation}
[-i\gamma^0\gamma\cdot\nabla+\gamma^0S(r)+V(r)-E_\alpha]u_\alpha(x)=0,\label{eq:D-eq}
\end{equation}
where $E_\alpha$ is the single quark energy. In general, the quark wavefunctions $u_\alpha(\vec x)$ may be expressed as
\begin{equation}
u_\alpha(\vec{x})=\left(\begin{array}{c}g_\alpha(r)\\\large{i\vec{\sigma}\cdot\hat{x}f_\alpha(r)}
\end{array}\right)\chi_s\chi_f\chi_c,
\end{equation}
where $\chi_s$, $\chi_f$ and $\chi_c$ are the spin, flavor and color quark wavefunctions, respectively.

\begin{figure}[t]
\flushleft
\includegraphics[width=0.157\textwidth]{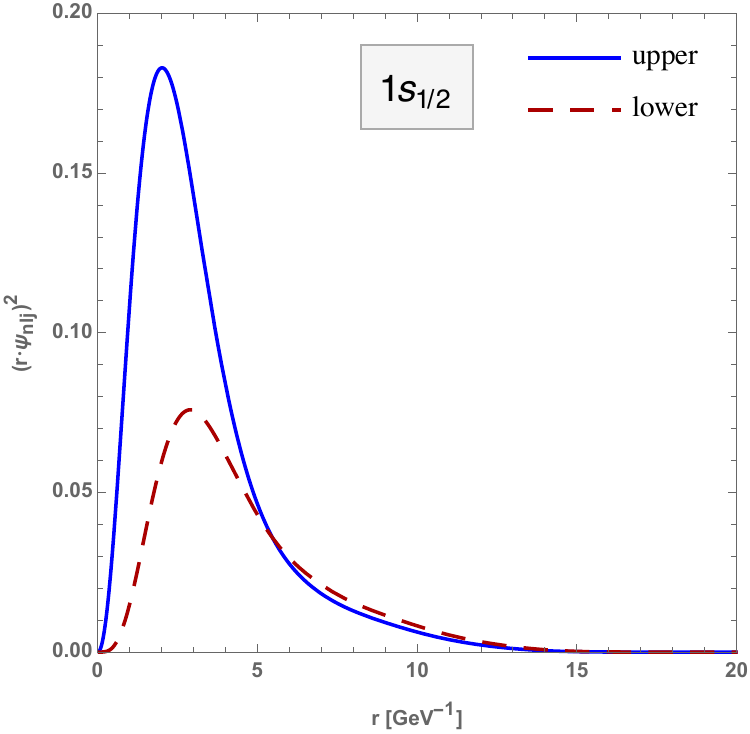}
\includegraphics[width=0.157\textwidth]{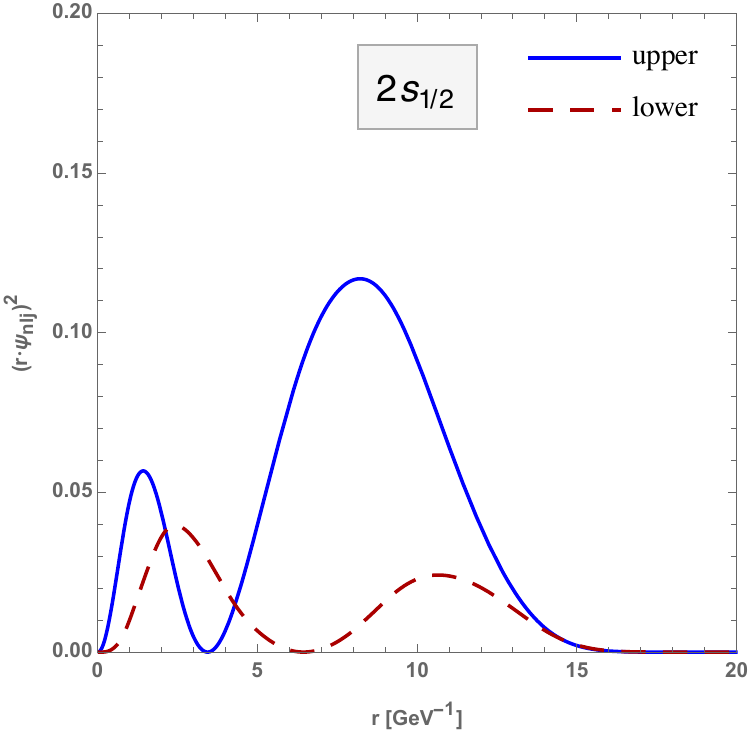}
\includegraphics[width=0.156\textwidth]{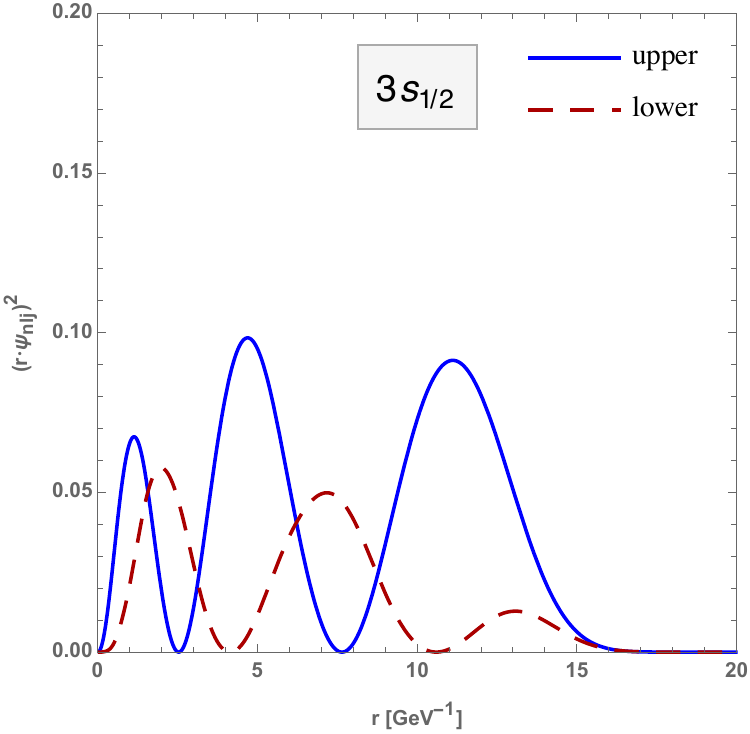}\\ \vspace{0.2cm}
\includegraphics[width=0.157\textwidth]{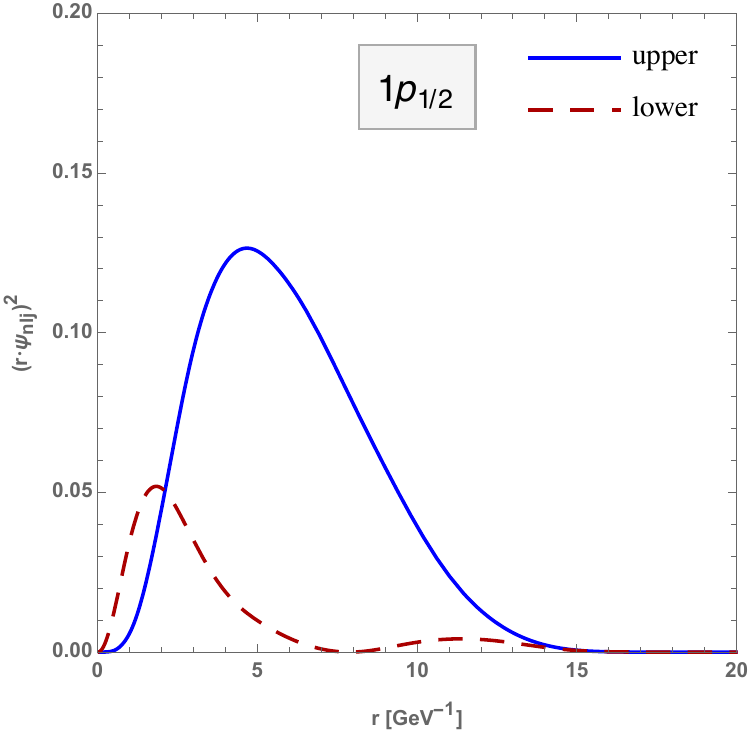}
\includegraphics[width=0.157\textwidth]{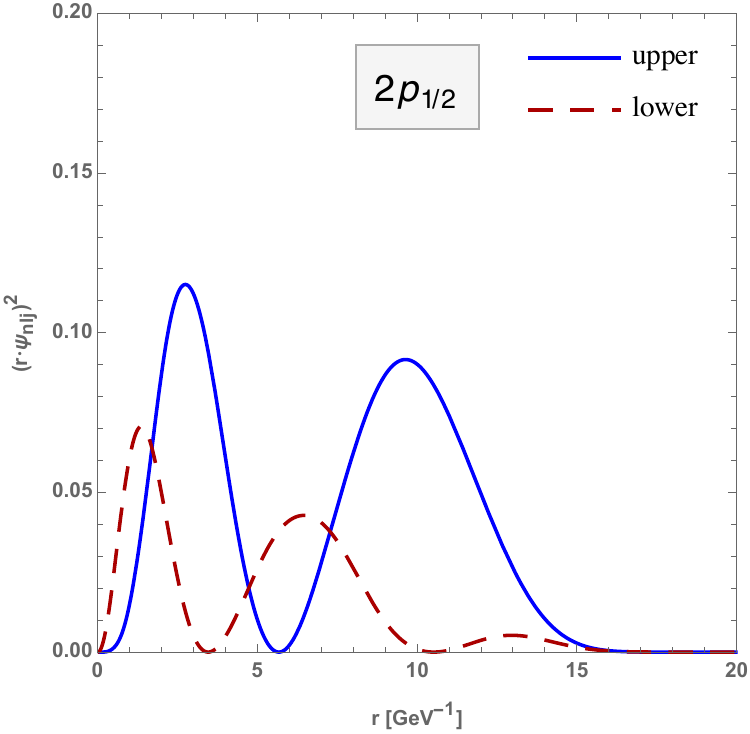}
\includegraphics[width=0.156\textwidth]{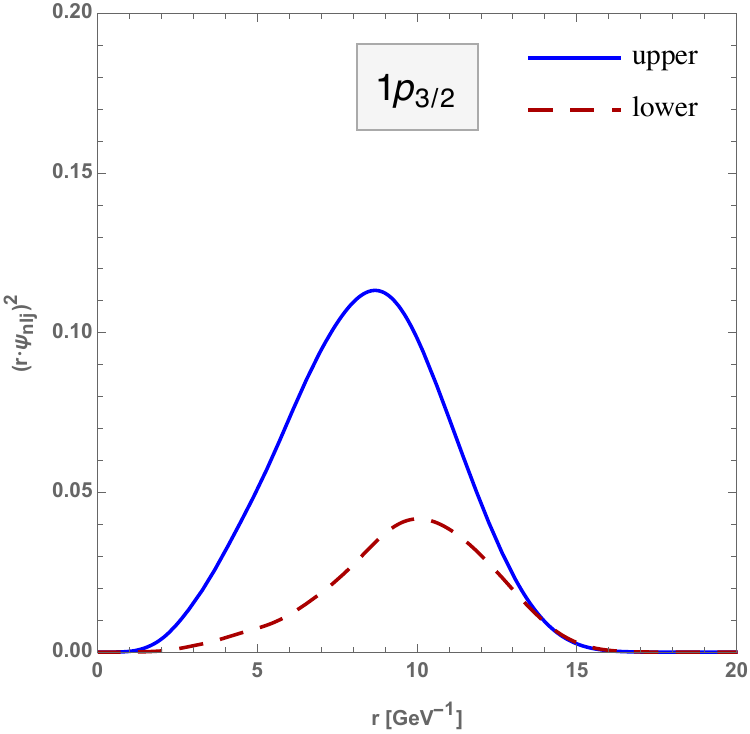}\\ \vspace{0.2cm}
\includegraphics[width=0.157\textwidth]{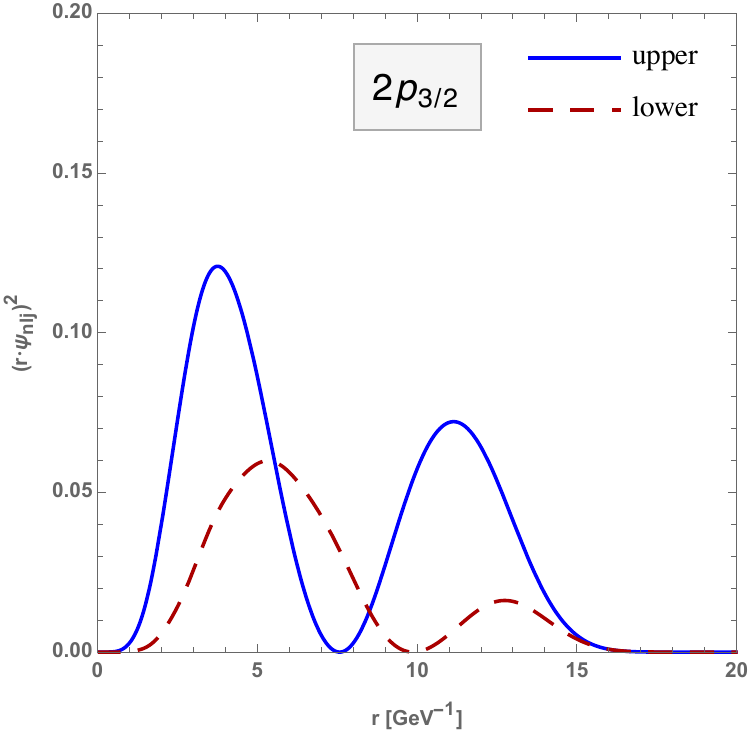}
\includegraphics[width=0.157\textwidth]{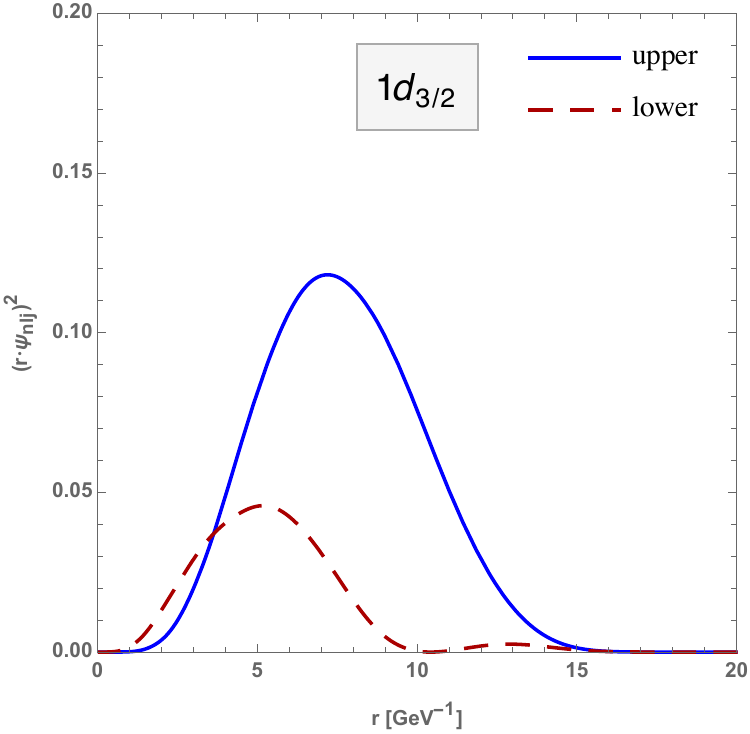}
\includegraphics[width=0.156\textwidth]{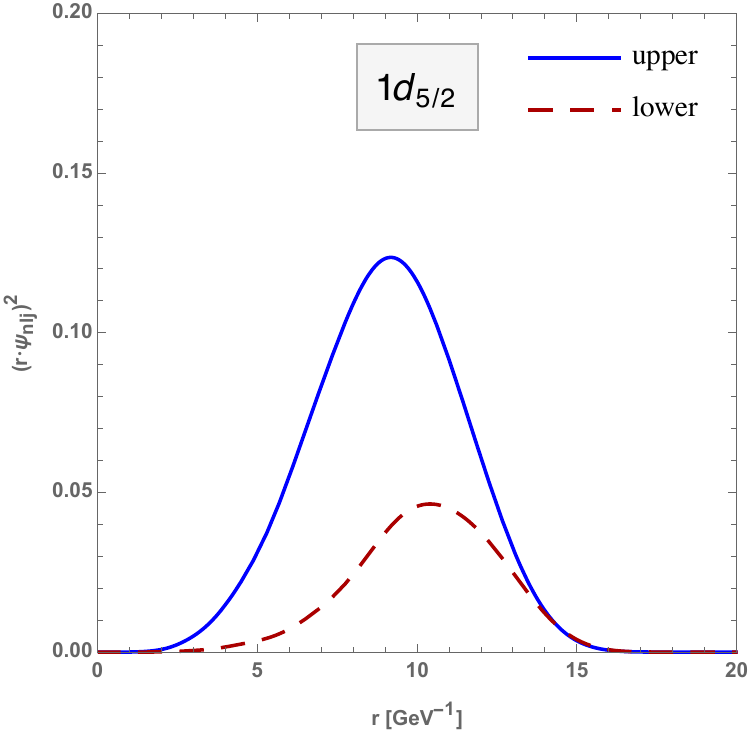}\\ \vspace{0.2cm}
\includegraphics[width=0.157\textwidth]{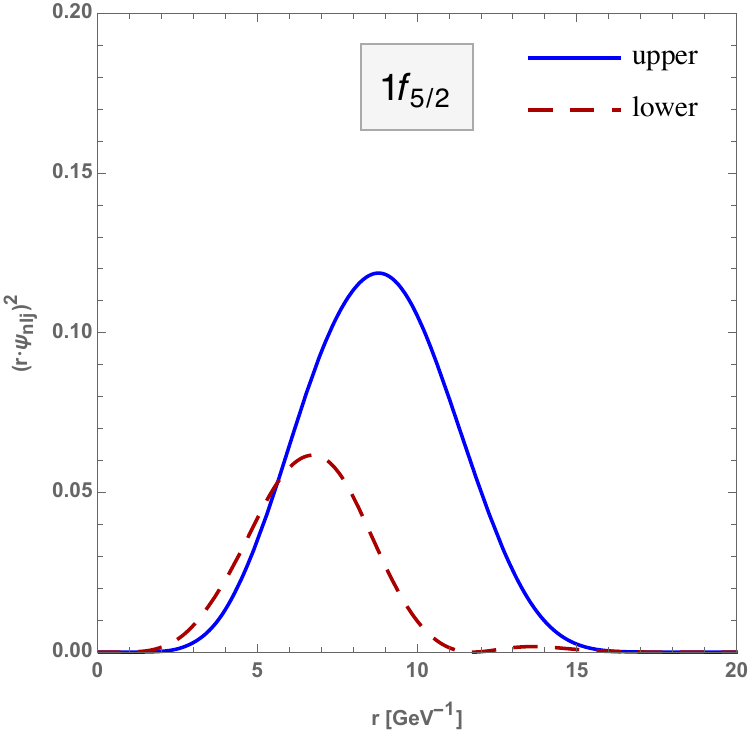}
\includegraphics[width=0.157\textwidth]{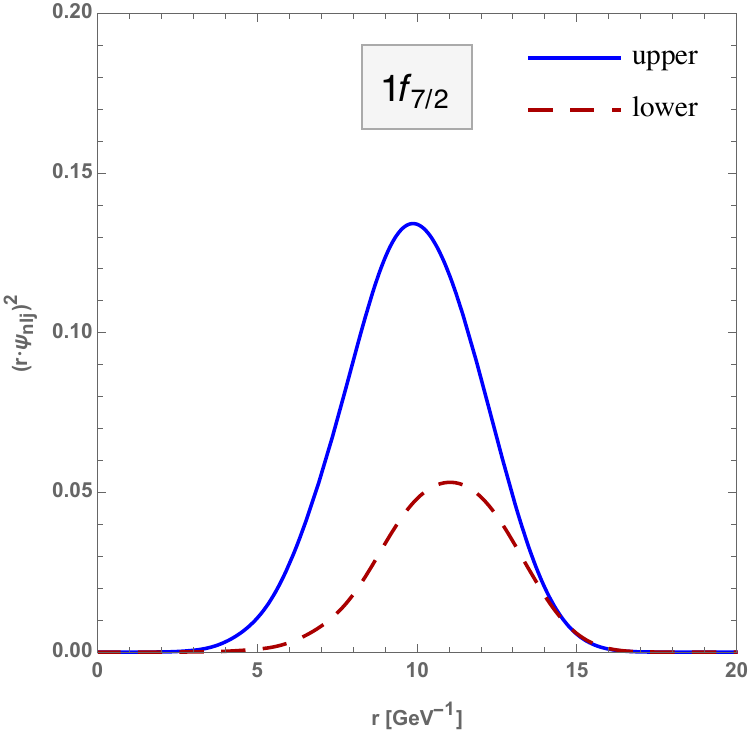}
\caption{The numerical results on the normalized radial wave functions of the valence quarks for the upper component $g(r)$ and the lower component $f(r)$ by solving Eq.~(\ref{eq:GEEP}) with the extracted PCQM potential $\Sigma(r)$ and $\Delta(r)$.}
\label{fig:QWF2}
\end{figure} 

In our previous works~\cite{Liu:2014,Liu:2015}, the ground state quark wavefunction $u_0(\vec x)$ has been determined by fitting the PCQM theoretical result of the proton charge form factor $G_E^p(Q^2)$ to the experimental data as shown in Fig.~\ref{fig:QWF}. The PCQM potentials $V(r)$ and $S(r)$ are derived by letting the ground state quark wavefunction $u_0(\vec x)$ satisfy the Dirac equation~(\ref{eq:D-eq}) and considering as boundary conditions the quark confinement and asymptotic freedom properties. In Fig.~\ref{fig:potential},  we present the extracted PCQM potentials $\Sigma(r)=V(r)+S(r)$ and $\Delta(r)=V(r)-S(r)$. It is clear that the potential $\Sigma(r)$ in the left panel of Fig.~\ref{fig:potential} shows a Cornell-like potential pattern. The potential $\Sigma(r)$ takes the form of the Coulomb potential at small $r$ but goes up quickly to infinite with $r$ increasing, which may be understood as the quark confinement. At the middle region of $r$, the potential $\Sigma(r)$ is nearly zero, which may indicate quarks are more or less free. The potential $\Delta(r)$ as shown in the right panel of Fig.~\ref{fig:potential} results in a mass wall.

Furthermore, the ground and excited quark wavefunctions could be derived numerically by solving Eq.~(\ref{eq:D-eq}) with the extracted PCQM potential using the Generalized Eigenvalue \& Eigenstate Problem method,
\begin{eqnarray}
&&H_{n'n}C_n=E_\alpha B_{n'n}C_n,\label{eq:GEEP}\\
&&H=\left(\begin{array}{cc} E_0+\Sigma(r) & -\frac{d}{dr}+\kappa/r\\ \frac{d}{dr}+\kappa/r & E_0+\Delta(r) \end{array}\right).\label{eq:H}
\end{eqnarray}
In the calculation, we expand the quark wavefunctions in the complete set of Sturmian functions $|S_{nl}(r)\rangle$~\cite{Liu:2014}. In Eq.~(\ref{eq:GEEP}), $E_\alpha$ and $C_n$ are the eigenvalues and eigenstates, respectively. $H_{n'n}$ are the matrix elements of operator $H$ of Eq.~(\ref{eq:H}) in the Sturmian basis, and $B_{n'n}=\langle S_{n'l}(r)|S_{nl}(r)\rangle$. $E_0$ is the ground state energy and has been determined as $E_0=0.524$ GeV in Ref.~\cite{Liu:2018}.

\begin{figure}[t!]
\begin{center}
\includegraphics[width=0.3\textwidth]{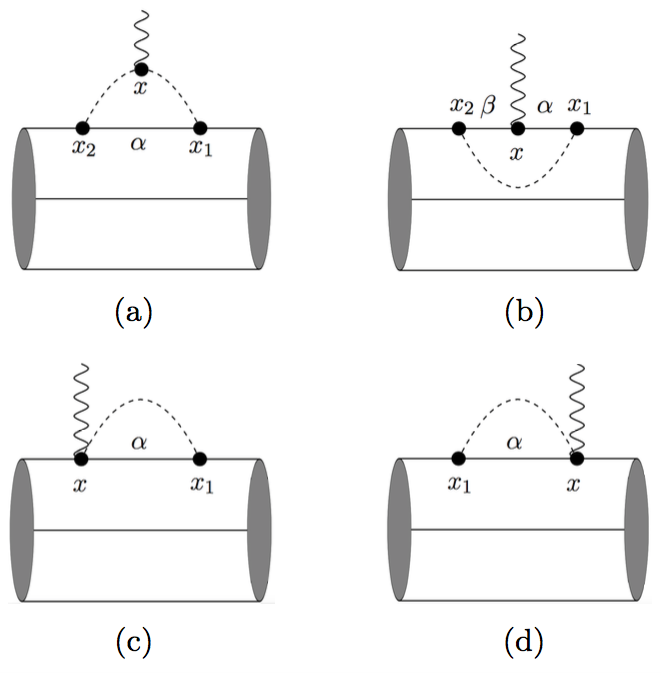}
\end{center}
\caption{Diagrams contributing to the charge form factor of neutron: meson cloud (a), vertex correction (b), self-energy I (c) and self-energy II (d).}\label{fig:GEnFF}
\end{figure} 

The results listed in Table~\ref{tab:energy} are the energy levels of a single quark, derived by solving Eq.~(\ref{eq:GEEP}) numerically with $E_0=0.524$ GeV. The radial wavefunctions of the valence quarks are presented in Fig.~\ref{fig:QWF2}, with the upper component $g(r)$ and the lower component $f(r)$. Note that in the calculation the quark wavefunctions are normalized according to $\int d^3\vec xu_\alpha^\dag(\vec x)u_\alpha(\vec x)=1.$ In this work, we restrict the energy levels $E_\alpha$ up to the low energy scalar $\Lambda=1$~GeV. Thus, the ground state $1s_{1/2}$ and the excited states $1p_{1/2}$, $1p_{3/2}$, $1d_{3/2}$, $1d_{5/2}$, $1f_{5/2}$, $1f_{7/2}$, $2s_{1/2}$, $2p_{1/2}$, $2p_{3/2}$ and $3s_{1/2}$ are included in the quark propagator to investigate the neutron charge form factor and charge radius in the PCQM.

\section{\label{sec:GEn}Neutron charge form factor in PCQM}

In the framework of the PCQM, the neutron charge form factors $G_E^n(Q^2)$ up to the one-loop order is given by
\begin{eqnarray}\label{eq:GEn}
\chi^\dag_{n_s'} \chi_{n_s}G^n_E(Q^2)&=&\,^n \langle{\phi_0}|\sum_{n=0}^2\frac{i^2}{2!}
\int\delta(t)d^4xd^4x_1 d^4x_2 e^{-i q\cdot
x}\nonumber \\
&& \times T[\mathcal{L}_I^W(x_1)\mathcal{L}_I^W(x_2)j^0(x)]|\phi_0\rangle_c^n,
\end{eqnarray}
where the state vector $|\phi_0\rangle^n$ corresponds to the unperturbed three-quark states projected onto the neutron state, which are constructed in the framework of the $SU(6)$ spin-flavor and $SU(3)$ color symmetry. The subscript \textit{c} in Eq.~(\ref{eq:GEn}) refers to contributions from connected graphs only. $\chi_{n_s}$ and $\chi^\dag_{n_{s'}}$ are the neutron spin wavefunctions in the initial and final states. $G_E^n(Q^2)$ is the neutron charge form factor. The charge current operator $j^0$ in Eq.~(\ref{eq:GEn}) is given by
\begin{eqnarray}\label{eq:c-current}
j^0&=& \bar\psi\gamma^0 {\cal Q}\psi+\bigg[f_{3ij}+\frac{f_{8ij}}{\sqrt{3}}\bigg]\Phi_i\partial_t\Phi_j\nonumber\\
&&+\bigg[f_{3ij}+\frac{f_{8ij}}{\sqrt{3}}\bigg]\frac{\Phi_j}{2F}\bar\psi\gamma^0\gamma^5\lambda_i\psi,
\end{eqnarray}
where ${\cal Q}=\textrm{diag}\{2/3, -1/3, -1/3\}$ is the quark charge matrix .

As discussed in our previous work [18], the 3q-core contributing to the $G_E^n(Q^2)$ vanishes due to the SU(6) structure, while the meson cloud or the loop diagrams dominate to the $G_E^n(Q^2)$ only. In Fig.~\ref{fig:GEnFF}, we present the Feynman diagrams contributing to the neutron charge form factor $G_E^n(Q^2)$ in accordance with the $\mathcal{L}_I^W(x)$ in Eq.~(\ref{eq:WT-int}) and the $j^0$ in Eq.~(\ref{eq:c-current}). For the sake of simplicity, calculations are restricted to the SU(2) flavor symmetry in this work, i.e. including $\pi$-meson cloud only, and the corresponding analytical expressions for the relevant diagrams are derived as follows:

\begin{widetext}       
\noindent(a) Meson cloud (MC) diagram:
\begin{eqnarray}
G_E^n(Q^2)\big|_{MC}&=&-\frac{1}{2(2\pi F)^2}\int_0^\infty dk k^2\int_{-1}^1 dx \frac{k^2+k Q x}{[\omega_\pi(k^2)+\Delta\mathcal{E}_\alpha][\omega_\pi(k_+^2)+\Delta\mathcal{E}_\alpha][\omega_\pi(k_+^2)+\omega_\pi(k^2)]}\nonumber\\
&&\times\bigg\{
F_{I\alpha}(k_+)F_{I\alpha}^\dag(k)+\Delta\mathcal{E}_\alpha \Big[F_{I\alpha}(k_+)F_{II\alpha}^\dag(k)+F_{II\alpha}(k_+)F_{I\alpha}^\dag(k)\Big]\nonumber\\
&&-\Big[\omega_\pi(k_+^2)\omega_\pi(k^2)+(\omega_\pi(k_+^2)+\omega_\pi(k^2))\Delta \mathcal{E}_\alpha\Big]F_{II\alpha}(k_+)F_{II\alpha}^\dag(k)\bigg\},
\end{eqnarray}
where $\omega_\pi(k^2)=\sqrt{M_\pi^2+k^2}$, $\Delta \mathcal{E}_\alpha=E_\alpha-E_0$ and
\begin{eqnarray}
F_{I\alpha}(k)&=&\int_0^\infty dr r^2\Big[g_0(r)g_\alpha(r)+f_0(r)f_\alpha(r)\cos2\theta\Big]\int_\Omega d\Omega e^{ikr\cos\theta}\mathcal{C}_\alpha Y_{l_\alpha0}(\theta,\phi),\label{eq:FI}\\
F_{II\alpha}(k)&=&\frac{i}{k}\int_0^\infty dr r^2\Big[g_0(r)f_\alpha(r)-g_\alpha(r)f_0(r)\Big]\int_\Omega d\Omega \cos\theta e^{ikr\cos\theta}\mathcal{C}_\alpha Y_{l_\alpha0}(\theta,\phi),\label{eq:FII}\\
k_\pm &=& \sqrt{k^2+Q^2\pm2k\sqrt{Q^2}x}.
\end{eqnarray}
The label $\alpha=(nl_\alpha jm)$ in the above equations characterizes the quark state. $\mathcal{C}_\alpha$ in Eq.~(\ref{eq:FI}) and~(\ref{eq:FII}) is the Clebsch-Gordan coefficients $C_\alpha=\langle l_\alpha0\frac{1}{2}\frac{1}{2}|j\frac{1}{2}\rangle$ and $Y_{l_\alpha0}(\theta,\phi)$ is the usual spherical harmonics with $l_\alpha$ being the orbital quantum numbers of the intermediate
states $\alpha$.\\

\noindent(b) Vertex correction (VC) diagram:
\begin{eqnarray}
G_E^n(Q^2)\big|_{VC}=\frac{1}{2(2\pi F)^2}\int_0^\infty dk k^4\frac{\mathcal{F}_{\pi\alpha\beta}(k)}{\omega_\pi(k^2)
[\omega_\pi(k^2)+\Delta\mathcal{E}_\alpha][\omega_\pi(k^2)+\Delta\mathcal{E}_\beta]} \cdot \mathcal{A}_{\alpha,\beta}(Q^2),
\end{eqnarray}
where
\begin{eqnarray}
\mathcal{F}_{\pi\alpha\beta}(k)&=&F_{I\alpha}(k)F_{I\beta}^\dag(k)
-\omega_\pi(k^2) F_{I\alpha}(k)F_{II\beta}^\dag(k)
-\omega_\pi(k^2) F_{II\alpha}(k)F_{I\beta}^\dag(k)
+\omega_\pi^2(k^2)F_{II\alpha}(k)F_{II\beta}^\dag(k),\\
\mathcal{A}_{\alpha,\beta}(Q^2)&=&\int_0^\infty dr r^2[g_\alpha(r)g_\beta(r)+f_\alpha(r)f_\beta(r)]\int_\Omega d \Omega e^{i Q r\cos\theta} \mathcal{C}_{\alpha,\beta}(\theta,\phi),
\end{eqnarray}
with
\begin{eqnarray}
\mathcal{C}_{\alpha,\beta}(\theta,\phi)&=&C_\alpha C_\beta Y_{l_\alpha0}(\theta,\phi)Y_{l_\beta0}(\theta,\phi)+D_\alpha D_\beta Y_{l_\alpha1}^\ast(\theta,\phi)Y_{l_\beta1}(\theta,\phi),\\
D_\alpha &=&\langle l_\alpha1\frac{1}{2}-\frac{1}{2}|j\frac{1}{2}\rangle.
\end{eqnarray}\\

\noindent(c) Self-energy I (SE I) diagram:
\begin{eqnarray}
G_E^n(Q^2)\big|_{SE:I}&=&-\frac{1}{4(2\pi F)^2}\int_0^\infty dk k^2\int_{-1}^1 dx\frac{k^2-kQx}{\omega_\pi(k^2)[\omega_\pi(k^2)+\Delta\mathcal{E}_\alpha]}\bigg[
\omega_\pi(k^2)F_{II\alpha}(k)F_{II\alpha}^\dag(k_-)-F_{I\alpha}(k)F_{II\alpha}^\dag(k_-)\bigg].\label{eq:SEI}\nonumber\\
\end{eqnarray}

\noindent(d) Self-energy II (SE II) diagram:
\begin{eqnarray}
G_E^n(Q^2)\big|_{SE:II}&=&-\frac{1}{4(2\pi F)^2}\int_0^\infty dk k^2\int_{-1}^1 dx\frac{k^2+kQx}{\omega_\pi(k^2)[\omega_\pi(k^2)+\Delta\mathcal{E}_\alpha]}\bigg[
\omega_\pi(k^2)F_{II\alpha}(k_+)F_{II\alpha}^\dag(k)-F_{II\alpha}(k_+)F_{I\alpha}^\dag(k)\bigg].\label{eq:SEII}\nonumber\\
\end{eqnarray}
\end{widetext}

The energy shifts $\Delta \mathcal{E}_\alpha$ in the above equations can be determined through the energy levels listed in Table~\ref{tab:energy}, and the corresponding quark wavefunctions of the ground and excited states are as shown in Fig.~\ref{fig:QWF2}. $G_E^n(Q^2)\big|_{SE:I}$ and $G_E^n(Q^2)\big|_{SE:II}$ respectively in Eq.~(\ref{eq:SEI}) and Eq.~(\ref{eq:SEII}) lead to the same results for the diagrams (c) and (d) of Fig.~\ref{fig:GEnFF} based on the T-symmetry, and then we label SE instead of the sum result of SE I and SE II diagrams in the following. It is noted that the $G_E^n(Q^2)$ is mainly attributed to the meson emission and re-absorption on the same quark in PCQM as shown in Fig.~\ref{fig:GEnFF}, while the diagram of meson exchange between two quarks is suppressed by the vertex function of $qq\pi$ system $F_{II0}(k)=0$ in Eq.~(\ref{eq:FII}).

\section{\label{sec:Results}Numerical results and discussion}

In this section we evaluate the $G_E^n(Q^2)$ and neutron charge radius with considering both the ground and excited quark wavefunctions which have been determined numerically in the section~\ref{sec:Potential&WF}. Note that there are no more parameters in the following numerical calculations on the neutron charge form factors.

\begin{table}[t!]
\caption{\label{tab:Netcharge}Numerical result for the net charge of neutron, which is the neutron charge form factor in zero-recoil $G_E^n(Q^2)|_{Q^2=0}$.}
\begin{tabular*}{\columnwidth}{@{\extracolsep{\fill}}lcc}
\hline
 \multicolumn{1}{l}{ } & Diagram & $G_E^n(0)$\\ 
\hline
                & MC & $-0.036$\\[2pt]
   Ground & VC & $\phantom{-}0.036$\\[2pt]
                & SE & $\phantom{-}0\phantom{-}\phantom{-}$\\[2pt]
                \cline{2-3}\\[-6pt]
                & MC & $-0.018$\\[2pt]
   Excited & VC & $\phantom{-}0.072$\\[2pt]
                & SE & $-0.054$\\[2pt]
                \cline{2-3}\\[-6pt]
   Total & MC+VC+SE & $\phantom{-}0\phantom{-}\phantom{-}$\\
\hline
\end{tabular*}
\end{table}

\begin{table}[b!]
\caption{\label{tab:rEn}Numerical results for the neutron charge radius $\langle r^2_E\rangle^n$ (in units of $\rm fm^2$). The experimental data are taken from Ref.~\cite{PDG:2016}, while the lattice QCD value is taken from~\cite{Alexandrou:2017,Shintani:2018}.}
\begin{tabular*}{\columnwidth}{@{\extracolsep{\fill}}lcccc}
\hline
   \multicolumn{1}{l}{ }&
   \multicolumn{1}{c}{$\langle r^2_E\rangle^n$}& Lattice~\cite{Alexandrou:2017} & Lattice~\cite{Shintani:2018} & \multicolumn{1}{c}{Exp.~\cite{PDG:2016}}\\[2pt]
\hline
   Ground & $-0.014$ & --- & --- & ---\\[2pt]
   Excited & $-0.058$ & --- & --- & ---\\[2pt]
   Total & $-0.072$ & $-0.038(34)$ & $-0.047(38)$ & $-0.116(2)$\\
\hline
\end{tabular*}
\end{table}

We first list in Table.~\ref{tab:Netcharge} the net charge of neutron, which is the neutron charge form factor in zero-recoil $G_E^n(Q^2)|_{Q^2=0}$. It is clear that the VC diagram contributes a positive value while both the MC and SE diagrams result in negative values. Ultimately, they counteract each other to let the net charge of neutron be zero exactly as the total result presented in Table~\ref{tab:Netcharge}. We also compile in Table~\ref{tab:rEn} the charge radius squared of neutron $\langle r^2_E\rangle^n$, which is derived by
\begin{equation}
\langle r^2_E\rangle^n=-6\frac{d}{dQ^2}G_E^n(Q^2)|_{Q^2=0}.
\end{equation}
The numerical results are separated into the contributions of the ground and excited states in the quark propagator. As listed in Table~\ref{tab:rEn}, we may point out that the excited quark state contributions are considerably influential in the neutron charge radius, and the total result of neutron charge radius increases fourfold when the excited states are included in the quark propagator although it is still rather smaller than the experimental value~\cite{PDG:2016}.

\begin{figure}[t!]
\begin{center}
\includegraphics[width=0.36\textwidth]{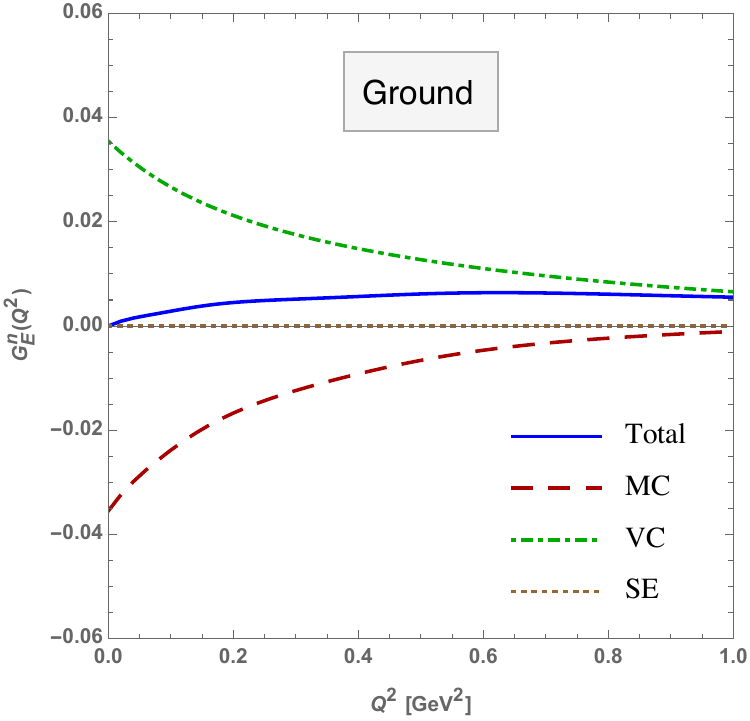}\hspace{0.5cm}
\caption{The individual contributions of the various diagrams of Fig.~\ref{fig:GEnFF} to the charge form factors of neutron attributed to the ground state quark propagator only.}
\label{fig:GEnGround}
\end{center}
\end{figure}

\begin{figure}[b!]
\begin{center}
\includegraphics[width=0.36\textwidth]{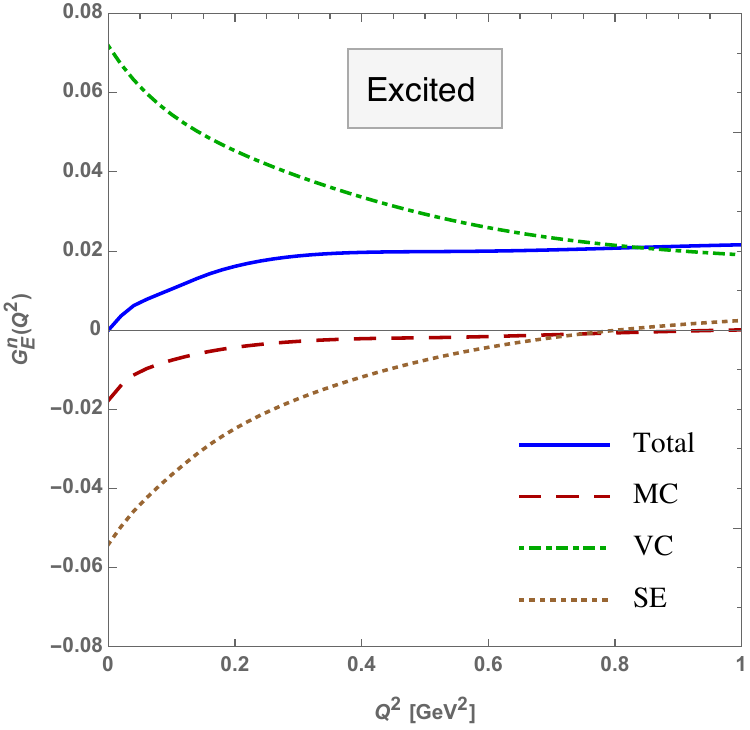}\hspace{0.5cm}
\caption{The individual contributions of the different diagrams of Fig.~\ref{fig:GEnFF} to the charge form factors of neutron when the excited states are included in the quark propagators.}
\label{fig:GEnExcited}
\end{center}
\end{figure}

In Fig.~\ref{fig:GEnGround}, we present the contribution of various processes as shown in Fig.~\ref{fig:GEnFF} to the $G_E^n(Q^2)$ in the low energy region $Q^2\leqslant1$ GeV$^2$ attributed to the ground state in quark propagator only. As the results shown in Fig~\ref{fig:GEnGround}, the contributions of the SE diagrams vanish exactly because of the vertex function of $qq\pi$ system $F_{II0}(k)\equiv0$ in Eqs.~(\ref{eq:SEI}) and~(\ref{eq:SEII}) for the ground state quark wavefunction. In this case, only the MC and VC diagrams contribute to the $G_E^n(Q^2)|_{Q^2=0}$ and counteract each other. We note that the total result based on the ground state quark wavefunction in Fig.~\ref{fig:GEnGround} leads to a small but nonvanishing $G_E^n(Q^2)$, which is the same as our previous finding of Ref.~\cite{Liu:2014}.

Further, we show in Fig.~\ref{fig:GEnExcited} the $Q^2$-dependence of the $G_E^n(Q^2)$ separated into the MC, VC and SE diagrams contributions, in which the intermediate excited quark states $1p_{1/2}$, $1p_{3/2}$, $1d_{3/2}$, $1d_{5/2}$, $1f_{5/2}$, $1f_{7/2}$, $2s_{1/2}$, $2p_{1/2}$, $2p_{3/2}$ and $3s_{1/2}$ as predetermined in Fig.~\ref{fig:QWF2} are included in the quark propagator. It is found that the VC and SE diagrams dominate the $G_E^n(Q^2)$, while the MC diagram contributes much less. The comparison between the results in Fig.~\ref{fig:GEnGround} and Fig.~\ref{fig:GEnExcited} indicates that the VC and SE diagrams with the excited state quark propagator contribute much more to the $G_E^n(Q^2)$ than the ones with the ground-state quark propagator.

\begin{figure}[t]
\begin{center}
\includegraphics[width=0.36\textwidth]{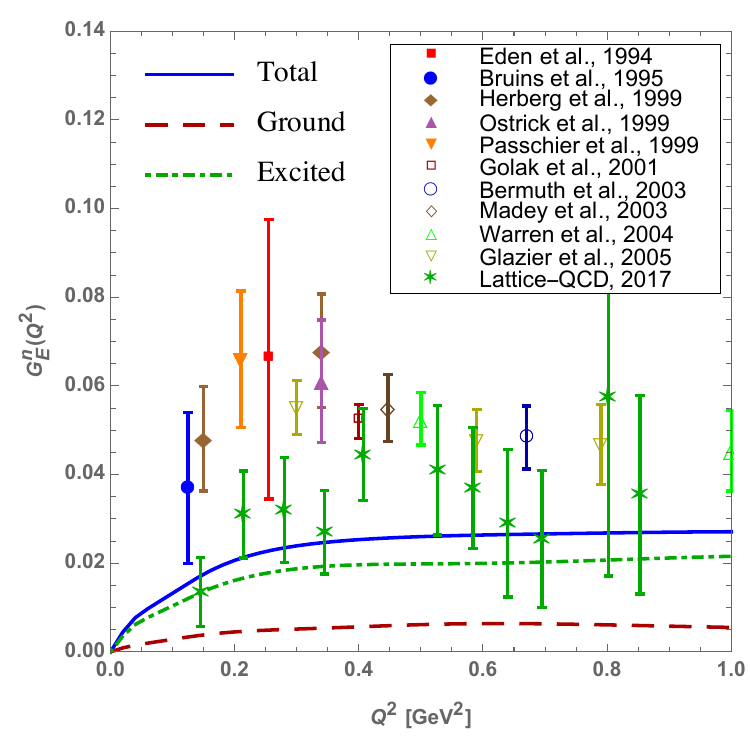}\hspace{0.5cm}
\caption{The PCQM results on neutron charge form factor $G_E^n(Q^2)$ considering both ground and excited states quark propagators. The experimental data on the neutron charge form factor are taken from Refs.~\cite{Eden:1994,Bruins:1995,Herberg:1999,Ostrick:1999,Passchier:1999,Golak:2001,Bermuth:2003,Madey:2003,Warren:2004,Glazier:2005}, and Lattice QCD results are from Ref.~\cite{Alexandrou:2017}}
\label{fig:GEnTotal}
\end{center}
\end{figure}

In Fig.~\ref{fig:GEnTotal}, we show the complete results on the $Q^2$-dependence of the $G_E^n(Q^2)$, including both the ground and excited states in the quark propagators, in comparison with the experimental data~\cite{Eden:1994,Bruins:1995,Herberg:1999,Ostrick:1999,Passchier:1999,Golak:2001,Bermuth:2003,Madey:2003,Warren:2004,Glazier:2005}. And we also plot the recent lattice QCD simulation results~\cite{Alexandrou:2017}, which simulations have been reproduced directly at physical pion mass of about 130 MeV. As the PCQM results show in Fig.~\ref{fig:GEnTotal}, the excited states quark propagators positively contribute to the $G_E^n(Q^2)$ and may be nearly fourfold over the one from ground-state quark propagator, and the total result of the $G_E^n(Q^2)$ is improved by including the excited states quark propagators as we expected. It is found that the total result of $G_E^n(Q^2)$ as shown in Fig.~\ref{fig:GEnTotal} is in good agreement with the lattice QCD values~\cite{Alexandrou:2017}, but it is still lower than experimental data. 

Obviously, the lattice QCD data of the $G_E^n(Q^2)$ shown in Fig.~\ref{fig:GEnTotal} underestimate the experimental ones, especially for $Q^2<0.2$ $GeV^2$. As discussed in Ref.~\cite{Alexandrou:2017}, the discrepancy between the lattice QCD results and the experimental data at small $Q^2$ may indicate that a larger spatial volume is required to fully develop the pion contributions. In addition, we note that the lattice QCD simulation in Ref.~\cite{Alexandrou:2017} could not count the contributions of the disconnected diagram, which is regarded as the isospin symmetry breaking term and positively shifts the lattice QCD evaluations of the $G_E^n(Q^2)$ as shown in Ref.~\cite{Alexandrou:2019}. Likewise, the PCQM is also restricted to the isospin symmetry limit, and hence the PCQM result of $G_E^n(Q^2)$ is consistent with the lattice QCD simulation but poor to the experimental data. Therefore, one may argue that it is necessary to consider isospin symmetry breaking for the $G_E^n(Q^2)$ in the PCQM.

Moreover, the center-of-mass correction plays an important role in relativistic quark models~\cite{Lu2:1998,Tursunov:2014,Dong:1999}. Ref.~\cite{Tursunov:2014} reveals that the nucleon mass is very sensitive to the center-of-mass effect, reduced by some 40\%.  In Ref.~\cite{Dong:1999}, the effects of the center-of-mass motion correction to the electroproduction transition of $N(1440)$ in the relativistic quark model have been studied and reported, while the theoretical results in Ref.~\cite{Lu2:1998} reveal that the center-of-mass correction increases the proton charge form factor by nearly 100\%. Thus, it is expected that the $G_E^n(Q^2)$ could be improved by considering the center-of-mass correction in the PCQM. 

\section{\label{sec:Summary}Summary and conclusions}

In this work, we have evaluated the charge form factor and charge radius of neutron in the PCQM with considering both ground ($1s_{1/2}$) and excited quark states ($1p_{1/2}$, $1p_{3/2}$, $1d_{3/2}$, $1d_{5/2}$, $1f_{5/2}$, $1f_{7/2}$, $2s_{1/2}$, $2p_{1/2}$, $2p_{3/2}$ and $3s_{1/2}$) in the quark propagator. The excited quark states are derived by solving the Dirac equation with the Cornell-like PCQM potential extracted in accordance with the predetermined ground state quark wavefunction. In summary, one may conclude that the neutron charge form factor $G_E^n(Q^2)$ is derived mainly from the meson emitted and reabsorbed on one quark in PCQM as shown in Fig.~\ref{fig:GEnFF}. 
The contributions of quark excitations reflected in the quark propagator are considerably influential in the $G_E^n(Q^2)$ as expected. The PCQM theoretical result of the $G_E^n(Q^2)$ is significantly improved, increased by nearly four times, by including the excited quark propagators, and it is in good agreement with the lattice QCD values with pion mass of about 130 MeV~\cite{Alexandrou:2017}. 

However, the PCQM theoretical result on the $G_E^n(Q^2)$ in this work is still lower than the experimental data. As we discussedin the previous section, the isospin symmetry breaking and the center-of-mass correction should be included in studying the $G_E^n(Q^2)$ in the PCQM, and it will be done in our future work.

\begin{acknowledgement}
This work is supported by the National Natural Science Foun- dation of China (Project No. 11547182), Suranaree University of Technology (SUT) and the Office of the Higher Education Commission (CHE) under the NRU project of Thailand. XL and AL acknowledge support from SUT-CHE-NRU (Contract No. FtR 09/2561). DS acknowledges support by Thailand re- search fund (TRF) under contract No. MRG5980255.
\end{acknowledgement}

\bibliographystyle{spphys}
\bibliography{Refs}

\end{document}